\input harvmac
\Title{\vbox{
\hbox{HUTP-00/A045}
\hbox{\tt hep-th/0011256}}}{An M-theory Flop as
 a Large N Duality}
\bigskip
\centerline{Michael Atiyah$^1$, Juan Maldacena$^{2,3}$ and
 Cumrun Vafa$^2$}
\bigskip
\centerline{$^1$ Mathematics Department}
\centerline{ University of
Edinburgh}
\centerline{Edinburgh, EH9 3JZ, UK}
\bigskip
\centerline{$^2$ Jefferson Physical Laboratory}
\centerline{Harvard University}
\centerline{Cambridge, MA 02138, USA}
\bigskip
\centerline{$^3$ Institute for Advanced Study}
\centerline{Princeton, NJ 08540}

\vskip .3in
We show how a recently proposed large $N$ duality
in the context of type IIA strings with ${\cal N}=1$
supersymmetry in 4 dimensions can be derived from
purely geometric considerations by embedding
type IIA strings in M-theory.  The phase structure of
M-theory on $G_2$ holonomy manifolds and an $S^3$ flop
are the key ingredients in this derivation.

\Date{November 2000}
\newsec{Introduction}
String propagation in the presence of branes
has been studied from many different angles, and has
been the intersection point of many fruitful ideas in the
context of dualities.  A beautiful example of it  is
in the context of the  AdS/CFT correspondence \ref\revar{
O.~Aharony, S.~S.~Gubser, J.~Maldacena, H.~Ooguri and Y.~Oz,
``Large N field theories, string theory and gravity,''
Phys.\ Rept.\  {\bf 323}, 183 (2000), hep-th/9905111.}.  
This idea is 
a refinement of the
statement that if we consider strings in the presence of branes,
in certain regime of parameters the system is better described
by strings propagating on the 
gravitational back reaction of light string modes to the presence
of branes.  This is an example of duality in the sense that
two different theories are continuously connected by a change
in parameters.  On the one end one has a geometry involving
branes and on the other extreme the geometry has been deformed and
the
branes have disappeared.

A recent example of a large $N$ duality in type IIA superstring,
 was proposed in \ref\va{C.~Vafa,
``Superstrings and topological strings at large N,''
hep-th/0008142.}\ based on embedding the large
$N$ Chern-Simons/topological gravity duality of 
\ref\gova{R.~Gopakumar and C.~Vafa,
``On the Gauge Theory/Geometry Correspondence,''
Adv.\ Theor.\ Math.\ Phys.\  {\bf 5}, 413 (1999), hep-th/9811131.}\
in type IIA superstrings.
The duality states that if we consider $N$ D6 branes wrapped on an
$S^3$ in the deformed 
 conifold geometry\foot{Mathematicians use the 
terminology ``quadric'' for the deformed 
conifold and ``quadric cone'' for a singular quadric.   The
small resolution of conifold is called 
the small resolution (blow up) of the
 quadric cone.} $T^*S^3$, then the same system
is equivalent to a type IIA geometry where the D-branes
have disappeared but where an $S^2$ has blown up so that
 the CY geometry is $O(-1)+O(-1)$ bundle
over ${\bf P}^1$. In other words, the topology is that 
of the so called ``small resolution'' of the conifold, where the
$S^2$ has finite size. 
  Moreover there are  $N$ units of 2-form field strength
flux through ${\bf P}^1$, and the complexified Kahler parameter $t$
of ${\bf P}^1$ is related to the volume $V$ of
the $S^3$ and the string coupling constant $g_s$ by
\eqn\fund{(e^t-1)^N= a\ exp(-V/g_s)}
Note that for large $V$ and when $Ng_{YM}^2={Ng_s\over V} \ll 1$, 
 the wrapped D-brane
description is good and the blown up description is 
bad as $t\rightarrow 0$,
and when $t\gg 0$ where the blown up ${\bf P}^1$ description
is good the $ V\rightarrow -\infty$ and the original wrapped
D-brane description is
bad.  
In this sense we only really have at most one good description
in each regime of parameters and the parameters being
related by \fund .  This situation is similar to other cases
one encounters in the context of large $N$ string dualities.
For example with $N$ D3 branes in $R^{10}$ if $Ng_{YM}^2=Ng_s \ll 1$ the
D-brane description ignoring the
gravitational backreaction is fine; when $Ng_s \gg 1$ the gravitationally
deformed background
description without the D-brane is the right description.

The main aim of this paper is to embed this type IIA duality in M-theory.
We find that the statement of this duality in the context of M-theory
translates to a simple geometric duality.  Turning it around, we can
derive the type IIA string duality of \va\ from its relation with
M-theory.

The organization of this paper is as follows.  In section
2 we review a perturbative string theory duality which is a good exercise
for the M-theory duality of interest.  In section 3 we discuss
M-theory in a 7 dimensional background with $G_2$ holonomy
and a simple geometric duality. In section 4 we reinterpret the
M-theory duality in the context of type IIA strings and obtain
the duality of \va .  In the appendix we discuss
some aspects of the $G_2$ holonomy metric. 

 The derivation of the large $N$ duality in this case
reinforces
the philosophy advocated in \gova\ and \va\ that large $N$
dualities in general correspond to  transitions in geometry.
It would be interesting to try to understand other
large $N$ dualities in the same spirit.

While we were completing this paper we received 
\ref\acha{ B. Acharya, 
`` On Realising N=1 Super Yang-Mills in M theory'',hep-th/0011089.}\
which has some overlap with this paper.

\newsec{A String theory duality}
Consider Type IIA superstrings propagating on
a non-compact CY background given by $O(-1)+O(-1)$
bundle over ${\bf P}^1$.  A powerful worldsheet description
of this sigma model is in terms of linear sigma model 
\ref\lsm{E.~Witten,
``Phases of N = 2 theories in two dimensions,''
Nucl.\ Phys.\  {\bf B403}, 159 (1993),
hep-th/9301042. } \
where one considers a $(2,2)$ supersymmetric $U(1)$ gauge theory
with four fields $(\Phi_1, \Phi_2, \Phi_3, \Phi_4)$ with charges
(+1,+1,-1,-1), with an FI term given by $r$ and a $U(1)$ $\theta$
angle. The low energy vacuum of this theory is characterized by
\eqn\vaco{{\cal V}:\qquad
|\Phi_1|^2+|\Phi_2|^2-|\Phi_3|^2-|\Phi_4|^2=r}
The actual vacuum of this theory is given by the gauge inequivalent
solutions to \vaco , i.e, one considers ${\cal V}/U(1)$.
This can  be naturally identified with $O(-1)+O(-1)$ bundle
over ${\bf P}^1$. If we take $r>0$ the ${\bf P}^1$ 
is identified as the locus $\Phi_3 =\Phi_4=0$, and the normal
directions are identified with $\Phi_3$ and $\Phi_4$.
This space is also called the ``small resolution'' of the conifold,
where the $S^2$ has finite size. 

An important aspect of this theory is its phase structure
and in particular what happens as $r\rightarrow 0$.  In fact
the natural phase structure for this theory is parameterized by
the {\it complex} parameter
$$t=r+i\theta.$$
It turns out that  both positive and negative $r$ make sense
and in fact are smoothly connected if we vary the $\theta$ because
the only singularity in the moduli space of this theory is
at the origin $t=0$.  {} From the equation \vaco\ one may naively
have thought that there should be some singularity at $r=0$ for
all $\theta$, but this is lifted by worldsheet instanton corrections.
Of course there is a simple reason why this had to happen.
The structure of $(2,2)$ supersymmetry leads to a naturally complex
moduli space and it
 does not allow any real
locus singularity in moduli space.

\subsec{A perturbative string duality}
 {}From the above description we notice a symmetry:  If we replace
$t\rightarrow -t$ we obtain the same geometry with the role
of the $(\Phi_1,\Phi_2)\leftrightarrow (\Phi_3 ,\Phi_4)$
exchanged.  Geometrically this is called a flop.  Even though
in the geometric setup there is a discontinuity at $r=0$,
and one just considers either $r>0$ or $r<0$, the situation in string
theory is different
because of the $\theta$ angle. Both regions are smoothly connected.
In particular if we start with $r\gg 0$ and do some computations
as a function of $t$, then analytic continuation of these quantities
by $t\rightarrow -t$ should yield the answers for the other side.
This in fact was directly checked at the level of instanton computations
on both sides \lsm \ref\asgm{P.S.~Aspinwall, B.R.~Greene and
D.R.~Morrison,``Multiple mirror manifolds and topology change in
string theory,'' Phys.Lett. {\bf B303} 249 (1993), hep-th/9301043.}.  
In particular the worldsheet instantons at genus 0
(with three points fixed on $S^2$) have a partition function
$$\partial^3_t F_0=C +{q\over 1-q}$$
where $q=e^{-t}$ (the constant term $C$ is somewhat ambiguous and
is related to the classical triple intersection of the 4-cycle dual to
${\bf P^1}$).  This makes sense as an instanton expansion when
$r \gg 0$.  However if we analytically continue this quantity
to $t\rightarrow -t$ we obtain in terms of ${\tilde q}=1/q$
$$-\partial^3_t F_0=1-C +{{\tilde q}\over 1-{\tilde q}}$$
which is the statement of the symmetry under $t\rightarrow -t$.
The shift in the constant term is reflecting the geometric
fact that under flop the classical triple intersection shifts by one.

Suppose however we consider a modified theory where we mod out
the $\Phi_i$'s by some discrete group $G$ which does not
necessarily act symmetrically
under the exchange $(\Phi_1 ,\Phi_2)\leftrightarrow (\Phi_3,\Phi_4)$.
  Let us call the resulting theory $Q_G[t]$,
exhibiting explicitly the dependence of the theory
$Q$ on the choice of the
group  $G$ as well as the (complexified) size of ${\bf P}^1$
given by $t$.  Now we ask what happens when we consider
$t\rightarrow -t$.  In this case in general we do not come back
to the same theory because of the asymmetric action
of $G$.  We obtain
\eqn\stdu{Q_G[-t]=Q_{G'}[t]}
where $G'$ is related to $G$ by conjugation with the element
exchanging the pairs
$$U:\qquad (\Phi_1,\Phi_2)\leftrightarrow (\Phi_3,\Phi_4)$$
$$G'=UGU^{-1}$$
To better appreciate the content of this duality statement
let us consider a simple example.  Consider the case where
$G$ is generated by the element
$$
(\Phi_1, \Phi_2 ,\Phi_3 ,\Phi_4)\rightarrow (\omega \Phi_1,
\omega^{-1} \Phi_2, \Phi_3, \Phi_4)$$
where $\omega$ is an $n$-th root of unity (this can
also be viewed as introducing an additional
$Z_n$ gauge group).  Geometrically this corresponds
to considering the following action on $O(-1)+O(-1)$
over ${\bf P}^1$:
$$(\zeta_1,\zeta_2,z)\rightarrow (\omega \zeta_1
,\omega \zeta_2,\omega^{-2}z)$$
where $z$ denotes the coordinate of ${\bf P}^1$
(say near the north pole) and
$\zeta_i$ denote the coordinates of the bundle 
$O(-1)+O(-1)$ over it.  Note in particular that
modding by this group leads to two orbifold
singularities, namely the north and
south poles at the origin of the $\zeta_i$.
  The group $G'$ on the
other hand is obtained by conjugating $G$ with $U$ and
is generated by
$$(\zeta_1,\zeta_2,z)\rightarrow (\omega \zeta_1 ,\omega^{-1} \zeta_2
,z)$$
This is a very different group action from $G$ and in particular
it leads to an $A_{n-1}$ singularity over ${\bf P}^1$.  Thus
changing $t\rightarrow -t$ has led to a totally new but
nevertheless ``dual''
theory.  Note that this duality is a perturbative
string
duality (i.e. can be understood genus by genus in
worldsheet expansion).

\newsec{An M-theory duality}
We will be interested in compactifications
of M-theory on $G_2$ holonomy manifolds (
see \ref\joyce{D.D.~Joyce,``Compact Riemannian 7-manifolds with
$G_2$ holonomy, I,II,'' J. Diff. Geom. {\bf 43}, 291 (1996) \semi
J. Diff. Geom. {\bf 43}, 329 (1996).}\ for construction
of some compact 7-manifolds with $G_2$ holonomy).  This leads to
${\cal N}=1$
supersymmetry in $d=4$.  In the compact case\foot{In the non-compact
case there could be additional deformations which change the
asymptotic behaviour of the metric at infinity.}, 
the number of moduli of the Ricci-flat metric
is given by the dimension of the third homology of the manifold, i.e.
$b_3$.  The point on the moduli space of Ricci-flat metrics  
can be characterized by the
volume of some basis for 3-cycles.  Physically we know that these should
correspond
to lowest components of chiral fields in ${\cal N}=1$ supersymmetry
multiplets.  Thus one again expects a complexification of the
volumes.  In fact this happens because there is a 3-form
gauge field $C$ in M-theory and its vev about each
3-cycle leads to the complexification of the volume elements.
Moreover the moduli space of M-theory compactifications are
given by analytic expressions and thus singularities occur
in {\it complex} codimension 1 and higher.  Thus there are
in particular no boundary walls in moduli space of M-theory.

Let us now come to the concrete example that we will be interested
in.  Consider the non-compact 7-manifold given by the
spin bundle over $S^3$.  This has the topology of $R^4\times S^3$
and admits a $G_2$ holonomy metric.
We will present the metric later in this section.  The topology of the
manifold can be viewed as 
$$(u_1^2+u_2^2+u_3^2+u_4^2)-(v_1^2+v_2^2+v_3^2+v_4^2)=V$$
where $u_i,v_i$ are real parameters. For $V>0$ the $S^3$ is identified
as the locus $v_i=0$, and $v_i$ correspond to the $R^4$ normal
directions over $S^3$.  Note that if we consider $V<0$
then the role of the $u$'s and $v$'s have been interchanged
and another $S^3$ blows up, corresponding to $u_i=0$.
 We can also write this in complex
notation by
$$(|z_1|^2+|z_2|^2)-(|z_3|^2+|z_4|^2)=V$$
where $z_i$ are complex variables. It should be emphasized 
that the $G_2$ holonomy manifold
does not admit a complex structure (it is odd dimensional) and
so there is no intrinsic meaning to writing the equation in terms
of complex coordinates, other than for simpler book keeping.
 We can also use a quaternionic
notation and write it as
$$|q_1|^2-|q_2|^2=V.$$
We can view the quaternion as $q_1=\sum_i u_i \sigma_i$ and
$q_2=\sum_i v_i \sigma_i$, where $\sigma_i$ denotes the
realization of quaternionic generators as $2\times 2$ matrices.
This way of writing it suggests that one can have an $SU(2)^{1,2}_{L,R}$
symmetries, which act on each quaternion,
as left and right multiplication by $SU(2)_{L,R}$.  This is the same,
as the $Spin(4)$ action on the $u$'s or on the $v$'s. 

A $G_2$ holonomy metric can be defined
on this manifold 
\ref\gibb{G.~W.~Gibbons, D.~N.~Page and C.~N.~Pope,
``Einstein Metrics On S**3 R**3 And R**4 Bundles,''
Commun.\ Math.\ Phys.\  {\bf 127}, 529 (1990).}\ref\brsal{R. Bryant 
and S. Salamon, `` On the 
construction of some complete metrics with exceptional holonomy'', 
Duke Math. J. {\bf 58} 829 (1989).}.  It is given as
 \eqn\gtwohol{
d s^2  = \alpha^2 dr^2  + \gamma^2 ({\tilde w}^a)^2 +
\beta^2 ( w^a -{ 1 \over 2} \tilde w^a)^2
}
with
\eqn\defof{
\alpha^{-2} = 1 - { a^3 \over r^3 } ~,~~~~~~~~ \beta^2 = { r^2 \over 9}
(  1 - { a^3 \over r^3 } )~,~~~~~ ~~ \gamma^2 = r^2/12
}
where $\tilde w^a$ and $ w^a$ are the left invariant one forms
on $\tilde S^3$ and $S^3$ respectively.  The two $S^3$'s are
associated to each of the two quaternions at fixed norm, and the
$r$ variable fills one of the $S^3$'s depending on the sign of
$V$.  In the form shown above $r \geq a$ and it fills
the $S^3$ associated with the left invariant one forms $w^a$,
while ${\tilde w}^a$ are associated with $\tilde S^3$ which is topologically
non-trivial.  Note that the volume of the $\tilde S^3$ is proportional
to $a^3$.
This  $G_2$ holonomy metric has an $[SU(2)]^3$ isometry group 
$$SU(2)^1_L\times
SU(2)^2_L\times [SU(2)^1_R \times SU(2)^2_R]_D$$
where $SU(2)^i_{L,R}$ denotes the $L,R$ multiplication
of $q_i$ by the $SU(2)$ group and
 $D$ denotes the diagonal subgroup. This is almost obvious
from the above presentation of the metric, where $SU(2)^i_L$ do not
act on the left-invariant forms and so do not modify the metric.
 The left-invariant 1-forms transform in the adjoint representation of
 the corresponding $SU(2)_R$. The diagonal combination of 
$SU(2)^i_R$ 
 leaves the metric invariant (the last
term in the metric is what requires choosing the diagonal
$SU(2)_R$ as the symmetry group). 
 To fix notation let us associate
$SU(2)^1$ with the $\tilde S^3$ sphere and $SU(2)^2$ with the 
contractible
$S^3$.
The space \gtwohol\ is asymptotic to a cone whose base has 
$\tilde S^3 \times S^3$ topology
$$ ds^2 \sim dr^2  + {r^2 \over 9} [ ( \tilde w^a)^2 + ( w^a)^2 - \tilde
w^a w^a ] 
$$
 As we will note later, when we discuss the connection to
type IIA string theory
we expect a one parameter family of $G_2$ holonomy
metric deformations which  breaks
 the $SU(2)^i_L$ to $U(1)$, for either  $i=1$ or $i=2$. 

As noted above, in the context of M-theory propagating in
this background the moduli space of the theory is parameterized
in addition to the volume $V$ of the $S^3$ by the flux of the
$C$-field through it.  Let us denote this complex combination
by $V_M=V+iC$.   Just as in the
case of string propagation on $O(-1)+O(-1)$ bundle over ${\bf P}^1$,
the phase structure of $M$-theory as a function of
 $V_M$ is expected to have a singularity at most only
at the origin and that turning on the $C$-field should smooth out
the singularity where $V=0$ in moduli space.  This follows from the number
of supersymmetries we are preserving in 4 dimensions and the fact that
$V_M$ is the lowest component of a chiral field.
A similar situation
(with twice as many supersymmetries) in the
context of M-theory and type IIA strings on Calabi-Yau threefolds
containing an $S^3$ was analyzed in 
\ref\oogv{H.~Ooguri and C.~Vafa,
``Summing up D-instantons,''
Phys.\ Rev.\ Lett.\  {\bf 77}, 3296 (1996), hep-th/9608079.}\
 where it was
 shown that Euclidean M2 brane instantons
modify the moduli space and in fact remove the singularity at $V=0$.
  Here also
we expect the same to be true, though we do not know how
to rigorously argue this.  At any rate we can argue, based
on supersymmetry alone, that a  singularity at most
will happen at isolated points in moduli space,
and for the theory at hand this means potentially only
at $V_M=0$ (in particular $V=0$ and $C\not= 0$ is not
a singular point).  This observation implies
that we can
continuously go from regions where the real part of $V_M$ is large
and positive to regions where it is large and negative without
encountering any singularity.\foot{For a discussion of
topology change in the context of $G_2$ holonomy manifolds
see the recent work \ref\pio{H. Partouche and B. Pioline,
``Rolling Among $G_2$ Vacua,'' hep-th/0011130.}.}
  Moreover it is clear that $V_M
\rightarrow -V_M$ is a flop, and otherwise gives rise to
an equivalent ``dual'' M-theory background.
Let us denote
M-theory in the presence of this
background by $Q[V_M]$.  Then we have
$$Q[V_M]=Q[-V_M]$$
Notice that for positive $V_M$, $S^3$ is contractible in 
the full geometry   and
$\tilde S^3$ is topologically nontrivial while the opposite 
is true of 
negative $V_M$.

Parallel to our discussion of the string duality in the
previous section, we can consider modding out by some group
actions which are isometries of the $G_2$ holonomy manifold.  As discussed
in the appendix this leads to manifolds (possibly with singularities) which
continue to have $G_2$ holonomy.
In
this way we obtain a statement of duality where
$$Q_G[-V_M]=Q_{G'}[V_M]$$
where $G=UG'U^{-1}$ and $U$ is the $Z_2$ outer automorphism
exchanging the $u$'s and the $v$'s, and acts on the $SU(2)$'s
as
$$U [SU(2)^{1,2}_{L,R}] U^{-1}=SU(2)^{2,1}_{L,R}$$
The special case we will be interested in is when
$G$ is generated by a $Z_N$ subgroup of $SU(2)^2_L$.
In complex coordinates we can view this transformation
as
$$g: \qquad
(z_1,z_2,z_3,z_4)\rightarrow (z_1,z_2, \omega z_3 ,\omega z_4)$$
Then $G'$ is generated by
$$g':\qquad (z_1,z_2,z_3,z_4)\rightarrow (\omega z_1,
\omega z_2, z_3, z_4)$$
and we have an M-theory duality
\eqn\given{Q_G[-V_M]=Q_{G'}[V_M]}
Let us consider $Q_G[V_M]$ when $V\gg 0$. In this case
the element $g$ acts with fixed point: The $\tilde S^3$ defined
by $z_3=z_4=0$.  Moreover the singularity is of the
type of $A_{N-1}$ singularity as the normal direction is
$R^4/Z_{N}$ in the usual action of $Z_N$ on $R^4$.
As is well known, this singularity in M-theory gives
rise to an $SU(N)$ gauge symmetry on the
singular locus.  In the present case, taking the number
of supersymmetries into account,  we have
an ${\cal N}=1$ supersymmetric $SU(N)$ Yang-Mills
gauge theory living on $\tilde S^3$ times the
Minkowski space.  On the other hand when we consider
$Q_{G'}[V_M]$ for $V\gg 0$, the $g'$ corresponds
to the $Z_N$ action with no fixed points (the would be
fixed point locus
$z_1=z_2=0$ is not on the manifold for $V\gg 0$).

\subsec{Gauge theoretic interpretation of the duality}
 {} From \given\ we see that M-theory in one background is
continuously connected with another one.  In particular
$Q_G[V_M]$ when $V \gg 0$ contains
as light excitations ${\cal N}=1$ Yang-Mills
sector
in 4 dimensions with the Yang-Mills coupling given by
$${1\over g_{YM}^2} +i\theta =V_M$$
where $g_{YM}^2$ should be viewed as the gauge coupling
at the Planck scale.  As we know the coupling
will start to run.  More precisely, the effective coupling
constant depends on the scale we probe.  Through the above
relation we see that $V_M$ itself should run and its value
will depend on which scale we measure it at.  In particular
 $V_M (\mu)$ should decrease logarithmically at infrared
as 
$$V_M(\mu )=V_M+{\rm const.} {\rm log}{ \mu\over M_{pl}}$$
This running should be induced by quantum effects
in the presence of the $Z_N$ singularity in measuring the volume of $S^3$,
 which is
at the singular locus.
For small $\mu$, we expect $V_M$ to become small.  In fact
if we trust the above formula we seem to  get a {\it negative}
volume $V_M$.  
Even though this is not allowed in the usual gauge
theory (negative $1/g_{YM}^2 $ naively does not make sense),
here we can make sense of negative $V_M$ as a flop. In fact
we are thus led to view the infrared behavior of the same theory
at negative and large values of $V\ll 0$.  However this theory
for negative $V$ is best viewed as the dual theory $Q_G[-V_M]
=Q_{G'}[V_M]$, in terms of which there is no singularity in geometry
and we obtain an ${\cal N}=1 $ theory in four dimensions with no sign
of $SU(N)$ gauge symmetry.  This is exactly what one expects
for a confining gauge theory.  Moreover we should see $N$ vacua.
This is also present here;  the $G'$ group corresponds
to modding out the $S^3$ by a $Z_N$. So the volume of the final
$S^3$ is smaller  by a factor of $V\rightarrow V/N$.  However we also have
to decide about the choice of the theta angle.  If we change
the theta angle by $2\pi k$ on $S^3$, which in the original
$Q_G[V_M]$ corresponds to not changing the theory at all,
as we go to negative $V_M$, it does give rise to a
change.  Namely
 quotienting the $S^3$ by a $Z_N$ gives rise to a fractional change in
the $C$-flux of the quotient theory by 
$2\pi k/N$.  Thus we obtain $N$ choices for the phase of the theory
in the infrared.  These are the $N$ vacua of ${\cal N}=1$ supersymmetric
$SU(N)$ Yang-Mills, and we have thus found a purely geometric
interpretation of them.  One can also identify the domain wall
of the ${\cal N}=1$ system with the $M5$ brane wrapped over $S^3/Z_N$.

\newsec{Re-interpretation of the M-theory duality in Type IIA
string}

We now view the same geometry from the type IIA perspective.
In order to do this we need to choose the ``11-th'' circle.
There are many ways to do this.  In order to connect this to the
duality of \va\ we identify  the 11-th direction 
with the fibers of the $U(1)$ sitting in $SU(2)^2_L$, where the
$Z_N$ that we modded out in the previous section,
is a subgroup of it.  In other words $Z_N \subset U(1) \subset
SU(2)_L^2 $.   We start with $Q_G[V_M]$
with $V \gg 0$.  Kaluza Klein reducing along this circle 
produces an $R^3$ fibration over $\tilde S^3$ with a singularity 
at the origin. 
%
%
The singularity
at the origin has the interpretation of a $D6$ brane,
before modding out by $Z_N$, or $N$ units of $D6$
branes after modding out.  Thus we expect this to correspond
to type IIA string theory  on the conifold background $T^* S^3$
with $N$ D6 branes wrapped around $S^3$.  
A more precise statement is the following.
Suppose we start with the deformed conifold before we put any branes
on it. 
In M-theory
we add an extra circle of constant radius so that we have 
a seven dimensional geometry which is the deformed conifold times a 
circle. Now we add $N$ D6 branes on ${\tilde S}^3$. 
At large distances
from the D6 branes, the presence
of the branes is signaled in IIA theory by the presence of a two
form field strength on the surrounding $S^2$. When we lift this up to M-theory
this means that the eleventh circle $S^1$ is non-trivially fibered over the 
$S^2$. In fact the total topology of this $S^1$ fibration over $S^2$ 
 ends up being that of $S^3/Z_N $. 
(Notice that for $ N=1$ we just have $S^3$). For very large $r$ we
expect that the dilaton will be constant, so that the size of the 
$S^1$ fiber is constant, while the size of $S^2\times {\tilde S}^3$ should
grow. For small $r$, on the other hand, we enter the near horizon
region of the six branes. The dilaton decreases as we approach the
``core'' of a six brane. In fact, the near horizon region of $N$
six branes in flat space lifts up in M-theory to an $A_{N-1} $ singularity
and 
the M-theory circle is just one of the angles on the three  sphere. 
So when we wrap this on $\tilde S^3$ we expect a geometry which 
is that of the region $r \sim a$ of \gtwohol . The asymptotics
of \gtwohol\ is not what we expect in the IIA situation since the 
radius of the M-theory circle continues to grow as $ r\to \infty$. 
In fact the geometry \gtwohol\ looks more like the infinite coupling
limit of the IIA geometry, where we take the limit in such a way 
to keep $V_M$ finite. 
In principle we expect to find a gravity solution in M-theory 
that describes more precisely the situation we expect in IIA theory
for finite string coupling constant.
That should be a deformation of the above $G_2$ holonomy metric
where the $SU(2)_L^2$ symmetry of 
\gtwohol\ is broken to $U(1)_L$ so that the circle can have constant
asymptotic size as $r \to \infty$. 
This deformation should exist  for both signs of $V_M$ which means,
in terms of \gtwohol , that we should also find a second 
deformation where $SU(2)_L^1$ is broken to $U(1)$. This would describe
the situation after the transition where, in IIA theory, 
 we have the 
small resolution of the conifold with $N$ units of $F_2$ flux. 
Again, we expect a solution where the string coupling asymptotes to a 
constant. 
Assuming this deformation of the metric exists, it is natural
to expect that under the flop $V_M\rightarrow -V_M$ we get
from one kind of deformation to the other.  In other words
the considerations of section 3 should also apply to this
deformed metric.  With this assumption we now rederive
the large $N$ type IIA duality, including the identification
of parameters on both sides in the geometric regime.

Let us denote the volume of the ${\tilde S}^3$ in the type IIA
setup by $V_A$, and consider $N$ D6 branes wrapped
over it. Then from the map between M-theory
parameters and type IIA parameters one deduces that
\eqn\oneo{V_M=V_A/g_s}
where $g_s$ is the type IIA coupling constant.  
  Now let us consider
the limit where $V_M \ll 0$.  In this case the theory is better
described by another M-theory background with group modding out
by $G'$, and where the volume of the $S^3$ is $-V_M$.  Again we
use the same 11-th direction for the circle fibration, which
means that we choose $G'$ to be a subgroup of the corresponding $U(1)$.
Now the fibration we get gives a geometry which has an $S^2$ and the
M-theory $S^3$ over it is a quotient of Hopf fibration by $Z_N$.
This means, in the type IIA terminology that we have $N$ units of
RR 2-form field strength through $S^2$.  Moreover the volume
of the (minimal) $S^2$ is given, changing the parameters from 
M-theory to type IIA, as
\eqn\twot{t=-V_M/N}
This relation is trustable for large $t$ where the supergravity
description of $M$-theory would be adequate.
Thus we see that we have a duality which in the type IIA description
corresponds to $N$ units of $D6$ branes
wrapping the $S^3$ of volume $V_A$ and on the other side
an $S^2$ with $N$ units of RR flux through it, with Kahler class
$t$ and with the relation (combining \oneo\ and \twot )
\eqn\mapp{t={-V_A\over N g_s}}
This agrees, in the limit of large $t$ with the result
obtained in \va :
\eqn\relsup{ (e^t-1)^N=a \ exp(-V_A/g_s) }
The modified relation \relsup\ includes the contribution
of worldsheet instantons, which is neglected in the 
identification at the large volume limit given in 
\mapp .  Note that in \va\ the parameter $t$ was
identified as the lowest component of the gaugino
chiral field $t=g_s Tr W^2=S$. Note that in the limit
of large $V$, which corresponds to small $t$ if we include
the instanton correction, we deduce a gaugino condensation
exactly as one would expect for the ${\cal N}=1$ Yang-Mills
theory, namely in the form $S^N=exp(-1/g_{YM}^2)$
(where the $g_{YM}$ is the Yang-Mills coupling constant at the string scale).

{} From the relation to 
M-theory it is natural that the worldsheet instantons
know about gaugino condensation.  This is similar to the
 worldsheet instantons
for the flopped geometry in the $O(-1)+O(-1)$ over ${\bf P}^1$.  Thus
the worldsheet instantons of the $O(-1)+O(-1)$
theory over ${\bf P}^1$ with $N$ units of flux, which in M-theory
correspond to superpotential corrections due to Euclidean
$M2$ brane instantons,
know, by analytic
continuation,
about the Euclidean $M2$ brane instantons of the flopped
geometry, which correspond to the usual gauge theory instantons.

Note that regardless of whether we find the new deformed $G_2$
holonomy solutions
or not, 
 we do not expect to find a geometry that truly decouples from the
bulk and which can be interpreted as a decoupled field theory. 
There is a limit where we expect a decoupled field theory, the
large $V_M$ limit, and the region very close to the
singularity, but it is not a limit where we expect a 
weakly coupled geometrical description.  This corresponds
as we saw above to $t \sim 0$.
 Though a weakly coupled
{\it string} desciption is expected for large $N$. 
The field theory we have been considering, therefore, has more 
degrees of freedom than pure ${\cal N}=1$ Yang-Mills. In particular, it 
has the parameter $V_M$ which is not a parameter in ${\cal  N}=1$ SYM. 

It is important to remark that when we talk about the resolved 
conifold of IIA theory with some units of flux, we are characterizing
the space in terms of its topology, but we do not have a complex 
manifold.  So in what sense is the consideration of topological
strings as used in \va\ relevant?  The answer turns out to be that
if we consider the BPS charge measured by the $M2$ brane,
when we integrate over the 11-th circle, it leads to a BPS charge
seen by the fundamental string, which corresponds to a symplectic
form on the quotient geometry, which topologically is $R^4 \times S^2$.
The symplectic structure induced from this reduction agrees with
the symplectic structure of $O(-1)+O(-1)$ bundle over ${\bf P}^1$,
as is shown in the appendix (the corresponding 
symplectic two form $k$ is obtained by integrating the
 G2 invariant three form $\Omega_3$
over the 11th circle, $k = \int_{S^1} \Omega_3$).
 This implies that we expect to obtain
the same results in consideration of topological strings for this reduction
of the 7-manifold.

\vglue 1cm
We have greatly benefited from discussions with 
R. Gopakumar, K. Hori, D. Joyce and A. Kovalev.

\vglue 1cm

The research of J.M.\ 
was supported in part by DOE grant DE-FGO2-91ER40654,
NSF grant PHY-9513835, the Sloan Foundation and the 
David and Lucile Packard Foundation.  The research of
C.V.\ was supported in part by NSF grant PHY-98-02709.

\appendix{A}{Aspects of $G_2$ Holonomy Metric}

A $G_2$ holonomy manifold is a seven dimensional manifold
whose holonomy group is the simple group $G_2$. It 
can be proven that in such manifolds there is a special harmonic
three form $\Omega$, $d \Omega = d * \Omega =0$, which is 
such that it locally determines the reduction of the 
holonomy group from Spin(7) to $G_2$. More precisely, at each point
the subgroup of GL(7) that leaves  $\Omega$ invariant  is $G_2$. 

It was shown in \brsal \gibb\
that the following metric (on $R^4\times S^3$)  has $G_2$ holonomy
 \eqn\gtwoholap{
d s^2  = \alpha^2 dr^2  + \gamma^2 ({\tilde w}^a)^2 +
\beta^2 ( w^a -{ 1 \over 2} \tilde w^a)^2
}
with
\eqn\defofap{
\alpha^{-2} = 1 - { a^3 \over r^3 } ~,~~~~~~~~ \beta^2 = { r^2 \over 9}
(  1 - { a^3 \over r^3 } )~,~~~~~ ~~ \gamma^2 = r^2/12
}
where $\tilde w^a$ and $w^a$ are left invariant one forms on 
two three spheres, which we denotes as $\tilde S^3 $ and $S^3$. 
We think of $S^3$ as the SU(2) group manifold. We can use the
following formulas: 
\eqn\defini{\eqalign{
g = & e^{ i \psi/2\sigma^3} e^{ i \theta/2 \sigma^1 }
 e^{ i \phi/2 \sigma^3 }
\cr
{ i \over 2} w^a_R \sigma^a  =& d g g^{-1} ~~~~~~~~~~~~
{i \over 2}  w^a_L \sigma^a  = g^{-1} d g 
\cr 
(w^1_R + i w^2_R ) = &e^{- i \psi} ( d \theta + i \sin \theta d\phi)
~~~~~~~~~~~ w^3_R = d \psi + \cos \theta d\phi 
\cr
(w^1_L + i w^2_L ) = &e^{+ i \phi} ( d \theta - i \sin \theta d\psi)
~~~~~~~~~~~ w^3_L = d \phi + \cos \theta d\psi 
\cr 
d w_R^a =& - {1 \over 2} \epsilon^{abc} w^b_R w^c_R
\cr
d w_L^a = & {1 \over 2} \epsilon^{abc} w^b_L w^c_L
}}
We can see from these definitions that the forms $w^a_L$ are invariant
under left multiplications of $g$, $ g \to h_L g$ while they transform
in the adjoint representation of $SU(2)$ under right multiplication
$g \to g h_R $. 
We can write the metric of the unit three sphere as
\eqn\metric{
ds^2 = { 1 \over 4} \sum_a  ( w_L^a )^2 = 
{ 1 \over 4} \sum_a  ( w_R^a )^2
}

We can easily check that the metric \gtwoholap\ has 
$SU(2)^3$ isometry. Two $SU(2)$s arise from left multiplication in
each of the $S^3$s while the third $SU(2)$ arises from right 
multiplication on both three spheres by the same group element. 
This last fact we can check by noticing that the index $a$, which 
transforms in the adjoint of $SU(2)$ is contracted in an $SU(2)$
 invariant
fashion. 

Finally we can write the explicit form of the three form
\eqn\simple{\eqalign{
\Omega  = & 
{  a^3 \over 12} 
{1 \over 6} \epsilon_{abc} 
\tilde w^a \tilde w^b \tilde w^c - { 1 \over 18} (r^3 -  a^3) 
\epsilon_{abc} (\tilde w^a w^b w^c - 
\tilde w^a \tilde w^b w^c ) + { r^2 \over 3} dr \tilde w^a w^a
\cr
= & {  a^3 \over 12} 
{1 \over 6} \epsilon_{abc} 
\tilde w^a \tilde w^b \tilde w^c  + d ({ r^3 -a^3 \over 9} 
 \tilde w^a w^a )
}}

{}From this expression we see that the $SU(2)^3$ isometry group of
the metric also leaves the three form invariant. In other words, these
symmetries leave the $G_2$ structure invariant. 

The metric \gtwoholap\ is asymptotic at large $r$ to 
\eqn\asymp{
ds^2 \sim 
  dr^2 + r^2 { 1 \over 9} [( {\tilde w}^a)^2 + ( {w}^a)^2 -  w^a 
{\tilde w}^a ]
}
The manifold on the right hand side of \asymp\ is a cone whose base
is topologically $\tilde S^3 \times S^3$. This cone has a singularity
at $r =0$. This singularity is eliminated in \gtwoholap\ by giving a
finite volume to one of the three spheres, $\tilde S^3$. We could 
similarly consider a situation where we give a finite volume to 
the other sphere $S^3$. In that case the manifold we obtain is 
just given by \gtwoholap\ with $w^a \leftrightarrow \tilde w^a$. 
These two manifolds are related by a flop. We see from the 
explicit expression of $\Omega$ that we can continuously go from one
to the other, passing through a singular manifold at the point
where both spheres have zero volume. 

If we qoutient this manifold by any subgroup of the isometry group
described above we will obtain again a $G_2$ manifold, since 
 the isometry group leaves the three form invariant. 
We will consider two different quotients. 

\subsec{ Singular quotient} 

In this quotient  we 
mod out by  $Z_N \subset U(1) \subset SU(2)^2_L$ which  acts
on the coordinates of $ S^3$. 
After modding out the metric \gtwoholap\ by $Z_N$ as above
we get a singular space. We get an $A_{N-1}$ singularity wrapped
over $\tilde S^3$. 
If we KK reduce over the circle associated to this $U(1)$ and 
we go to type IIA theory the singularity looks like
the singularity we have in the near horizon region of N sixbranes
wrapped on $\tilde S^3$. The normal bundle of this $S^3$ in IIA
theory is the same as
 what we have when we wrap branes on the $S^3 $ of a deformed 
conifold. In more mathematical notation, it is  $T^* \tilde S^3$. 
Notice that in this IIA description, there is a singularity at
the position of the branes even for ${\cal N}=1$\foot{
Mathematically the  point is that
when we quotient $R^4 = C^2$ by the circle (acting as  complex scalars)
the resulting space can be  naturally identifed with $R^3$ topologically
but not differentiably -  the identification being singular at the
origin.  This singularity in type IIA is interpreted as a D6 brane.}. 
As usual, in the near horizon region of six branes in IIA theory,
the dilaton is varying and it is approaching zero at the 
core of the sixbranes \ref\itzhaki{
N.~Itzhaki, J.~M.~Maldacena, J.~Sonnenschein and S.~Yankielowicz,
``Supergravity and the large N limit of theories with sixteen  
supercharges,''
Phys.\ Rev.\  {\bf D58}, 046004 (1998), hep-th/9802042.}.
 The string metric in the directions
along the six brane is also shrinking as we approach the core, but it
does so in such a way that $V_M = V_A/g_s$ is constant and equal to
the volume of $\tilde S^3$ in M-theory.  The above remarks about the
near horizon region of a six brane apply close to the singularity at 
$r =a$, for the region $r-a \ll a $.  For large $N$ there is a large
region where the IIA description is valid \itzhaki . 
As we go futher away from the singularity the geometry becomes 
more and more strongly coupled and the 11 dimensional description 
becomes better. For large $r$ the dilaton goes to infinity. In principle
there should be another solution where the dilaton goes to a constant.

\subsec{ Non-singular quotient}

If we choose $Z_N \subset U(1) \subset SU(2)^1$. This group acts 
as left multiplication on $\tilde S^3$. Since the volume of 
this three sphere is nowhere vanishing we conclude that the 
quotient is non-singular.  We can further Kaluza Klein reduce
the metric along this circle. This produces a non-singular IIA metric
on a space which has the topology of the small resolution of
the conifold and with $N$ units of two form flux on $S^2$. 
If $N$ is very large, then this IIA geometry is weakly coupled at
the origin. When we start moving out in the radial direction the
string coupling starts increasing and it becomes infinite asymptotically.
In principle there should be another solution where the string coupling
does not diverge. 
If we integrate the three form \simple\ on this $S^1$ we find
the symplectic form on the small resolution of the conifold. 
It should be noted that, though we get the same two form as the Kahler
form of the local CY, the 
IIA geometry with the flux is not Kahler. 

Let us see this more explicitly. The circle in question is 
parametrized by $\tilde \psi$. In order to do the KK reduction 
it is convenient 
to define the new one forms $\hat w^a$ and
the vector $n^a$ as follows
\eqn\defn{
\tilde w^a_L = \hat w^a + n^a d \tilde \psi 
}
Notice that $n^a n^a =1$.
We have extracted the dependence on $\tilde \psi$ in order to KK
reduce. This splitting preserves  $SU(2)^1_R$ but breaks
$SU(2)^1_L$
to $U(1)_L$.

The IIA dilaton, metric and one form RR potential are 
\eqn\tendkk{\eqalign{
e^{ 4 \phi/3} = & { 1 \over N^2} [ \gamma^2 + {\beta^2 \over 4} ] 
\cr
ds^2_{str} = & e^{ 2 \phi/3} \left[ dx_4^2  +  \alpha^2 dr^2 
- e^{ 4 \phi/3}  A_1^2 + \gamma^2 (\hat w^a )^2  + 
\beta^2 ( w^a - { 1\over 2 } \hat w^a)^2 \right]
\cr
A_1   = & N \left[ 
\hat w^a n^a  -  { 
 \beta^2  \over 2 ( \gamma^2 + \beta^2/4) } w^a n^a \right]
}}
where $A_1$ is the RR one form potential. 

Similarly we can integrate the three form over $S^1$ and we obtain
\eqn\twof{
J = { a^3 \over 12} { 1\over 2} \epsilon_{abc} n^a \hat w^b \hat w^c 
  - ({r^3 \over 18 } - { a^3 
\over 18} ) \epsilon_{abc} ( n^a w^b w^c -2 n^a w^b \hat w^c ) 
 - { r^2 \over 3} dr n^a w^a
}
which is closed since $\Omega $ was closed.

We can further simplify these expressions by doing a coordinate 
transformation that amounts to switching from the left invariant
one forms to the right invariant one forms in $\tilde S^3$. 
This translates into the following replacements
\eqn\repla{
\hat w^a  \to  \check w^a_R ~~~~~~~~~~~~~ n^a \to \check n^a_R ~~~~~~~~~~
w^a \to w^a + \check w^a_R 
}
Where now $\check w^a_R , \check n^a_R$ are defined through 
$$ 
\tilde w^a_R|_{\tilde \psi =0}  = \check w^a_R + \check 
n^a_R d\tilde \psi ~.
$$
In these variables 
$$ \check  n^a_R = \delta^a_3, ~~~~ \check w^1 = d\tilde \theta ~,~~~~
\check w^2 = \sin\tilde \theta d\tilde \phi ~,~~~~
\check w^3 = \cos \tilde \theta
d\tilde \phi ~. 
$$
In these variables the two form \twof\  becomes 
\eqn\twoform{
J = { a^3 \over 12} \sin \tilde \theta d \tilde \theta d\tilde \phi 
  - d ( { r^3 - a^3 \over 9} (w^3 + \cos \tilde \theta d \tilde \phi) 
)}
This agrees with the two form of the small resolution of the conifold
which is 
\eqn\smallres{
k = t \sin \tilde \theta d \tilde \theta d\tilde \phi 
  - d ( h( \rho) ( w^3 + \cos \tilde \theta d \tilde \phi )
)}
where $h(\rho)$ is some function of the radial coordinate.
these two expressions coincide if we identify $h(\rho) = (r^3 -a^3)/9$
which just amounts to a reparametrization of the radial coordinate. 

We can also write the IIA metric \tendkk\ and
the RR 1-form potential in a more explicit form
\eqn\tendexp{\eqalign{
ds^2_{str} = & e^{ 2 \phi/3} \left[ dx_4^2  +  \alpha^2 dr^2 
 + \gamma^2 ( d \tilde \theta^2 + \sin^2 \tilde \theta d \tilde \phi^2 )
  + \right. \cr
 & \left. 
\beta^2[ ( w^1 + { 1\over 2} d \tilde \theta)^2 +
( w^2 + { 1\over 2} \sin \tilde \theta d \tilde \phi )^2 + 
( w^3 + \cos \tilde \theta d \tilde \phi )^2 ]  - \right. \cr
 & \left. 
{  \beta^4  \over 4 ( \gamma^2 + \beta^2/4) } 
 (  w^3 + \cos \tilde \theta d \tilde \phi )^2 \right]
\cr
A_1   = & N \left[ \cos \tilde \theta d \tilde \phi   - 
 { \beta^2  \over 2 ( \gamma^2 + \beta^2/4) }
( w^3 + \cos \tilde \theta d \tilde \phi )  \right]
}}

\listrefs
\end